\documentclass[3p,sort&compress]{elsarticle}
\usepackage{lineno}
\usepackage{amssymb}
\usepackage{amsmath}
\usepackage{multirow}
\usepackage{adjustbox}
\usepackage{epsfig} 
\usepackage{threeparttable}
\usepackage{color}
\usepackage{widetext}
\usepackage{xcolor}
\usepackage{float}
\usepackage{times}
\usepackage[colorlinks,citecolor=blue,linkcolor=red,anchorcolor=blue,filecolor=blue,urlcolor=blue]{hyperref}

\journal{Nuclear Physics A}
\begin{document}

\begin{frontmatter}
\title{Superheavy Nuclei and the Changing Face of Nuclear Magicity}
\author[address1]{Jeet Amrit Pattnaik}
\ead{jeetamritboudh@gmail.com}
\cortext[correspondingauthor]{Corresponding author}
\author[address2]{Santosh Kumar}
\author[address2]{S. K. Singh}
\author[address1]{R. N. Panda}
\author[address3]{M. Bhuyan}
\author[address1]{S. K. Patra}

\address[address1]{Department of Physics, Siksha $'O'$ Anusandhan, Deemed to be University, Bhubaneswar-751030, India}
\address[address2]{Patliputra University, Patna-800 020, India}
\address[address3]{Institute of Physics, Sachivalya Marg, Bhubaneswar-751005, India}
\date{today}
\begin{abstract}
Using a relativistic mean field formalism, we analyzed the magic number sequence for finite nuclei in the superheavy valley. The result for the IOPB-I parameter set is compared with the well-known NL3 force. The magic numbers obtained from IOPB-I and NL3 interactions are found to be similar. Analysing the single-particle levels and the number of nucleons occupied in it, we find the close shell sequence as 2, 8, 18, 34, 50, 58, 80, 82, 92, 114, 120, 120, 138, 164, 172, 184 and 198 for the $^{318}{120}$ mass region. Again, with a careful inspection, we noticed large shell gaps at nucleon numbers 2, 8, 18, 34, 50, 58, 80, 92, 120, 138, 164, 172, 184, and 198, which may be considered as the magic number sequence for the superheavy nuclei. This change may be due to the shape change of the nuclear potential as compared to the stability valley. 
\end{abstract}
\end{frontmatter}
\section{Introduction} \label{intro} 
Similar to the atomic shell closure, the nuclear shell closure is an old, but important terminology in nuclear physics \cite{kkumar}. Historically, the nuclear shell closure was quite mysterious, so these numbers are termed as magic numbers. That means, the nucleon numbers 2, 8, 20, 28, 50, 82 for protons and 2, 8, 20, 28, 50, 82, and 126 for neutrons are well known magic numbers in the $\beta-$stability valley \cite{mayer49}. These numbers are well reproduced by solving the Schr\"odinger equation with a central plus spin-orbit potential. In other words, we find large shell gaps in the single-particle energy levels in the filling of the nucleons at the magic numbers. However, the same magic number sequence is reproduced directly by solving the Dirac Lagrangian, as the spin-orbit interaction is inbuilt. By solving the relativistic Lagrangian, we incorporate directly the actual nuclear interaction, including the central, tensor and spin-orbit potentials. Again, with a proper set of parameters, which are capable of reproducing the finite nuclear properties for the entire range of the mass table as well as the nuclear matter properties for sub- and supra-saturation densities, one can extend the calculations to the superheavy regions for the prediction of shell closure and magic number sequence. \\

Now, the question arises, whether the same shell closure appears in the superheavy and drip-line regions?  Are there substantial shell gaps as compared to standard shell gaps in these finite nuclei area? Whether some of the shell gaps are too small, so that we will not consider them as magic numbers. Chou et. al. \cite{chou95} reported that in the region of exotic nuclei, the traditional magic numbers lose their close shell nature, but bear a different sequence. Patra and group \cite{gupta97a,gupta97b} undertook extensive studies and predicted different sequences of double shell close for the exotic and superheavy regions \cite{sil04,sahoo20}. They also identified proton number $Z = 120$ as the next magic number in the superheavy valley with neutron number N = 184 as the corresponding combination. The main motivation of the present analysis is to answer some of the above questions, which generally come to the mind of a nuclear physicist. Here, we use the effective field theory motivated relativistic mean field  (E-RMF) Lagrangian to find out the single-particle energy levels in a spherically symmetric coordinate space calculations \cite{estal01a,estal01b}. The well-known IOPB-I parameter set \cite{kumar18} is used to evaluate the finite nuclear properties. To validate the model, we have also compared the results with the celebrated NL3 force \cite{lala97}.

A large number of predictions are available for the nuclear magic number. Many of them indicate the Z = 114 could be the next proton magic number beyond Z = 82 \cite{patyk91,nix92,nix94,gre69,nil69}. Although the N = 126 is the known largest neutron magic number, the proton magic number does not coincide with it, because of the huge Coulomb repulsion \cite{patra04} and probably Z = 114 is the next close shell configuration. However, after the discovery of Fl (Z = 114) and the E-RMF calculations, despite the shell closure at Z = 114, the shell gap is very small for this configuration, which does not qualify to be a magic number \cite{bhumpla}. The search for the next magic number beyond Z = 82 remains unsettled, prompting further investigation. After a through investigation, we find Z = 120 as the next magic number in the superheavy valley and the corresponding neutron magic number is located at N = 184 \cite{bhumpla,gupta97a,gupta97b,rutz97,shakeb12,pat23pramana}, which synthesization can be confirmed in near future \cite{gate24,roberto23}. There are several theoretical criteria formulated to judge the shell filling as magic number, such as a large shell gap in the closure shell, zero of the pairing gap in a self-consistent calculation, a sudden fall in the two-nucleons separation energy in an isotopic/isotonic chain, etc \cite{pat21scr}. A detailed description can be found in Ref. \cite{bhumpla}. To understand the reason behind the formation of magic number, however, we have to analyze the nuclear potential, because it determines the distribution of nucleons inside the nucleus. In the present article, our motivation is to understand the filling of both protons and neutrons for the heavy nuclei region in the framework of a relativistic mean field analysis \cite{estal01a,estal01b}.
\section{The Relativistic Mean Field Formalism}
\label{theory}
The relativistic mean field formalism is designed as the interaction of the nucleons with an iso-scalar-scalar ($\sigma$) meson, an iso-scalar-vector ($\omega$) meson, an iso-vector-vector ($\rho$) and iso-vector-scalar ($\delta$) mesons for asymmetric system and of course the Coulomb interaction takes care of the charge of the protons. The direct interaction of the nucleons through these mesons exchange and the self- and cross-couplings of these bosons are also considered in the effective field theory motivated relativistic mean field (E-RMF) Lagrangian \cite{frun96,frun97,estal01a,estal01b}. A detailed formalism and the construction of the force parameter on the E-RMF Lagrangian can be found in Refs. \cite{kumar18,kuma17,frun96,frun97}. From the last few decades, the E-RMF formalism has gained popularity for its spectacular success in finite nuclei, nuclear matter, and studies of nuclear stars, including the properties of Gravitational Waves physics \cite{kumar18,kuma17,mali2018}. The E-RMF energy density functional is given by \cite{kumar18}:
\begin{widetext}
\begin{eqnarray}
{\cal E}({r}) & = &  \sum_\alpha \varphi_\alpha^\dagger({r})
\Bigg\{ -i \mbox{\boldmath$\alpha$} \!\cdot\! \mbox{\boldmath$\nabla$}
+ \beta \left[M - \Phi (r) - \tau_3 D(r)\right] + W({r})
\nonumber \\[3mm]
& &
+ \frac{1}{2}\tau_3 R({r})
+ \frac{1+\tau_3}{2} A ({r})
- \frac{i \beta\mbox{\boldmath$\alpha$}}{2M}\!\cdot\!
  \left (f_\omega \mbox{\boldmath$\nabla$} W({r})
  + \frac{1}{2}f_\rho\tau_3 \mbox{\boldmath$\nabla$} R({r}) \right)
  \Bigg\}
\nonumber \\[3mm]
& & \null   
  \varphi_\alpha (r)
  + \left ( \frac{1}{2}
  + \frac{\kappa_3}{3!}\frac{\Phi({r})}{M}
  + \frac{\kappa_4}{4!}\frac{\Phi^2({r})}{M^2}\right )
   \frac{m_s^2}{g_s^2} \Phi^2({r})
-  \frac{\zeta_0}{4!} \frac{1}{ g_\omega^2 } W^4 ({r})
\nonumber \\[3mm]
 & &  \null
+ \frac{1}{2g_s^2}\left( 1 +
\alpha_1\frac{\Phi({r})}{M}\right)
\left(\mbox{\boldmath $\nabla$}\Phi({r})\right)^2
\nonumber \\[3mm]
 & &  \null
 - \frac{1}{2g_\omega^2}\left( 1 +\alpha_2\frac{\Phi({r})}{M}\right)
 \left( \mbox{\boldmath $\nabla$} W({r})  \right)^2
 - \frac{1}{2}\left(1 + \eta_1 \frac{\Phi({r})}{M} +
 \frac{\eta_2}{2} \frac{\Phi^2 ({r})}{M^2} \right)
 \nonumber \\[3mm]
 & &  \null 
  \frac{m_\omega^2}{g_\omega^2} W^2 ({r})
   - \frac{1}{2e^2} \left( \mbox{\boldmath $\nabla$} A({r})\right)^2
   - \frac{1}{2g_\rho^2} \left( \mbox{\boldmath $\nabla$} R({r})\right)^2
   - \frac{1}{2} \left( 1 + \eta_\rho \frac{\Phi({r})}{M} \right)
 \nonumber \\[3mm]
 & &  \null 
   \frac{m_\rho^2}{g_\rho^2} R^2({r})
   -\Lambda_{\omega}\left(R^{2}(r)\times W^{2}(r)\right)
   +\frac{1}{2 g_{\delta}^{2}}\left( \mbox{\boldmath $\nabla$} D({r})\right)^2
   +\frac{1}{2}\frac{ {  m_{\delta}}^2}{g_{\delta}^{2}}\left(D^{2}(r)\right)\;.
\label{eq1}
\end{eqnarray}
\end{widetext}
The fields $\Phi$, $W$, $R$ and $D$ are redefined as $\Phi = g_s\sigma$, $W = g_\omega \omega$, $R$ = g$_\rho\vec{\rho}$, and $D = g_\delta\delta$ by multiplying their respective coupling constants $g_\sigma$, $g_\omega$, $g_\rho$ and $g_\delta$. The masses of the $\sigma$, $\omega$, $\rho$ and $\delta$ mesons are  $m_\sigma$, $m_\omega$, $m_\rho$ and $m_\delta$, respectively. The well-known Coulomb coupling constant $\frac{e^2}{4\pi}$ is taken in the calculations. The symbols used in Eq. (\ref{eq1}) carry their usual meaning and can be found in Refs. \cite{kuma17,kumar18}. Using the Euler-Lagrangian relation for the equation of motion for finite nuclei, we get a set of coupled differential equations for each meson and nucleon from Eq. (\ref{eq1}), which are solved self-consistently in an iterative method. We calculate the single-particle energy, neutrons and protons distribution radii, nuclear potential, spin-orbit interaction for the desired finite nuclei.

\section{RESULTS AND DISCUSSIONS}
\label{results}
\begin{figure}
\centering
\includegraphics[width=0.8\columnwidth]{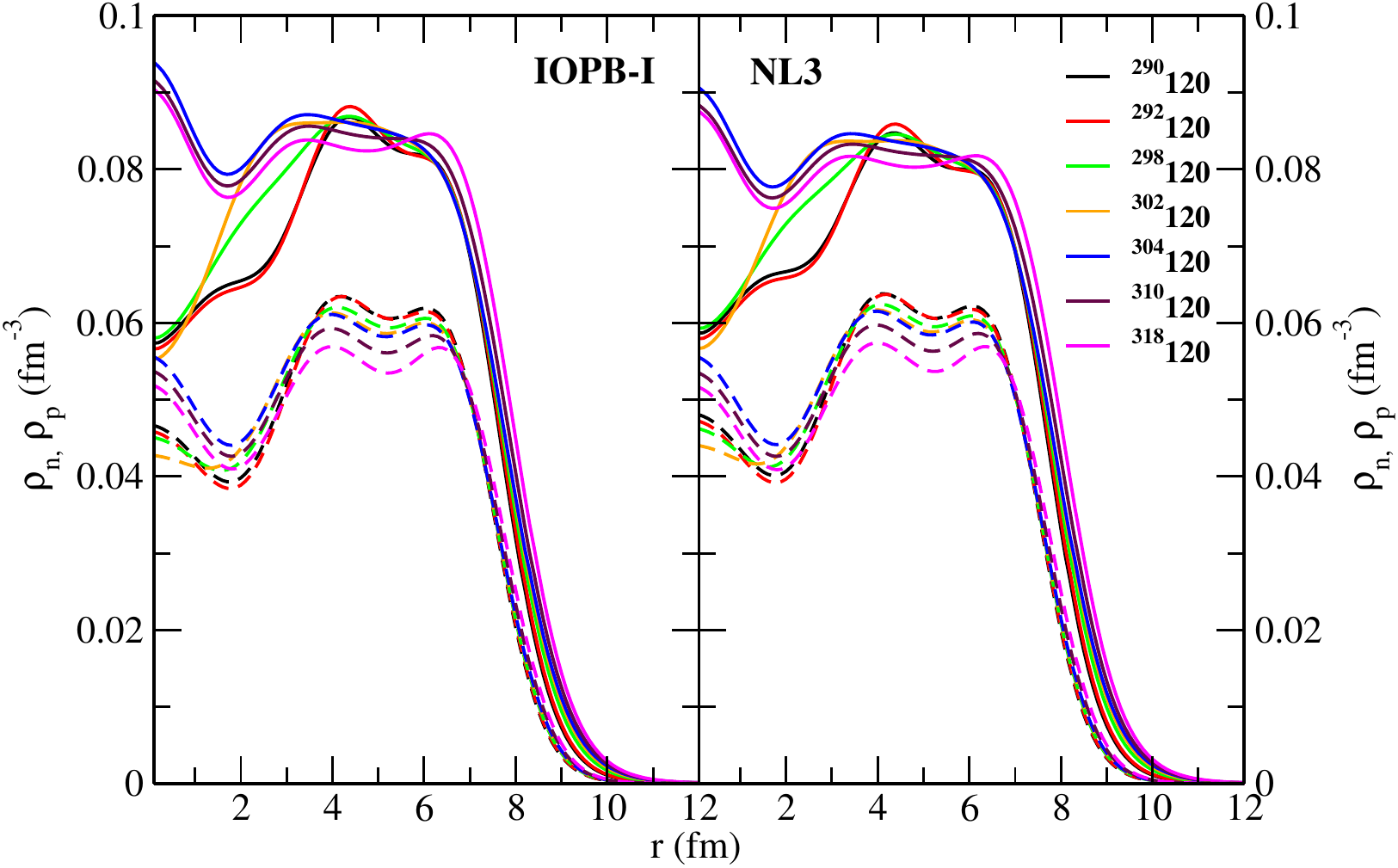}
\caption{The neutrons and protons densities distribution for $^{290,292,298,302,304,310,318}$120 with IOPB-I and NL3 parameter sets. The solid lines for the neutron distributions and the dashed lines for the proton density distributions. } 
\label{fig1}
\end{figure}

\begin{figure}
\centering
\includegraphics[width=0.8\columnwidth]{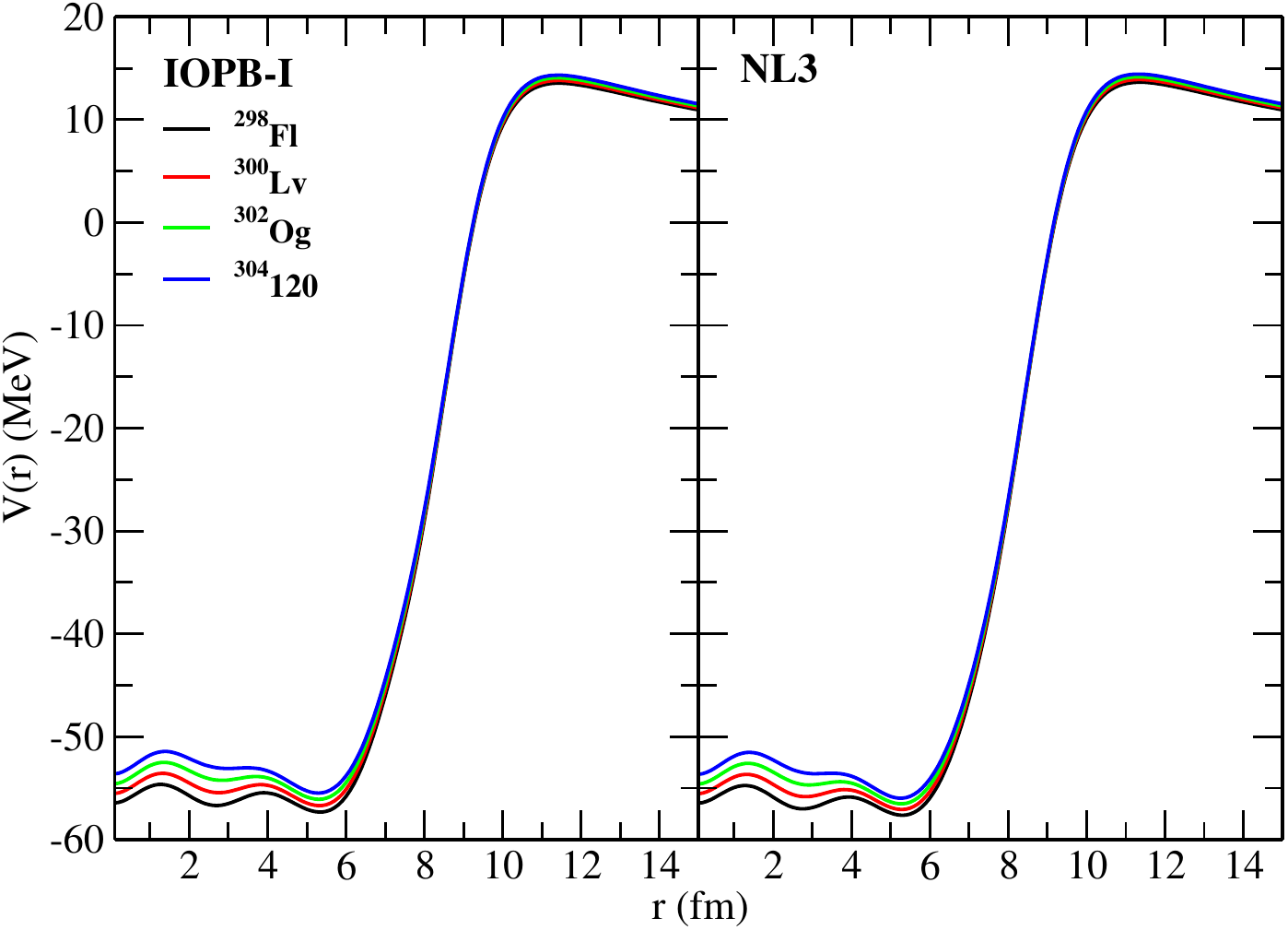}
\caption{The nuclear potential $V(r)$  for $^{298}$Fl, $^{300}$Lv, $^{302}$Og and $^{304}$120 with IOPB-I and NL3 parameter sets. } 
\label{fig2}
\end{figure}

\begin{figure}
\centering
\includegraphics[width=0.8\columnwidth]{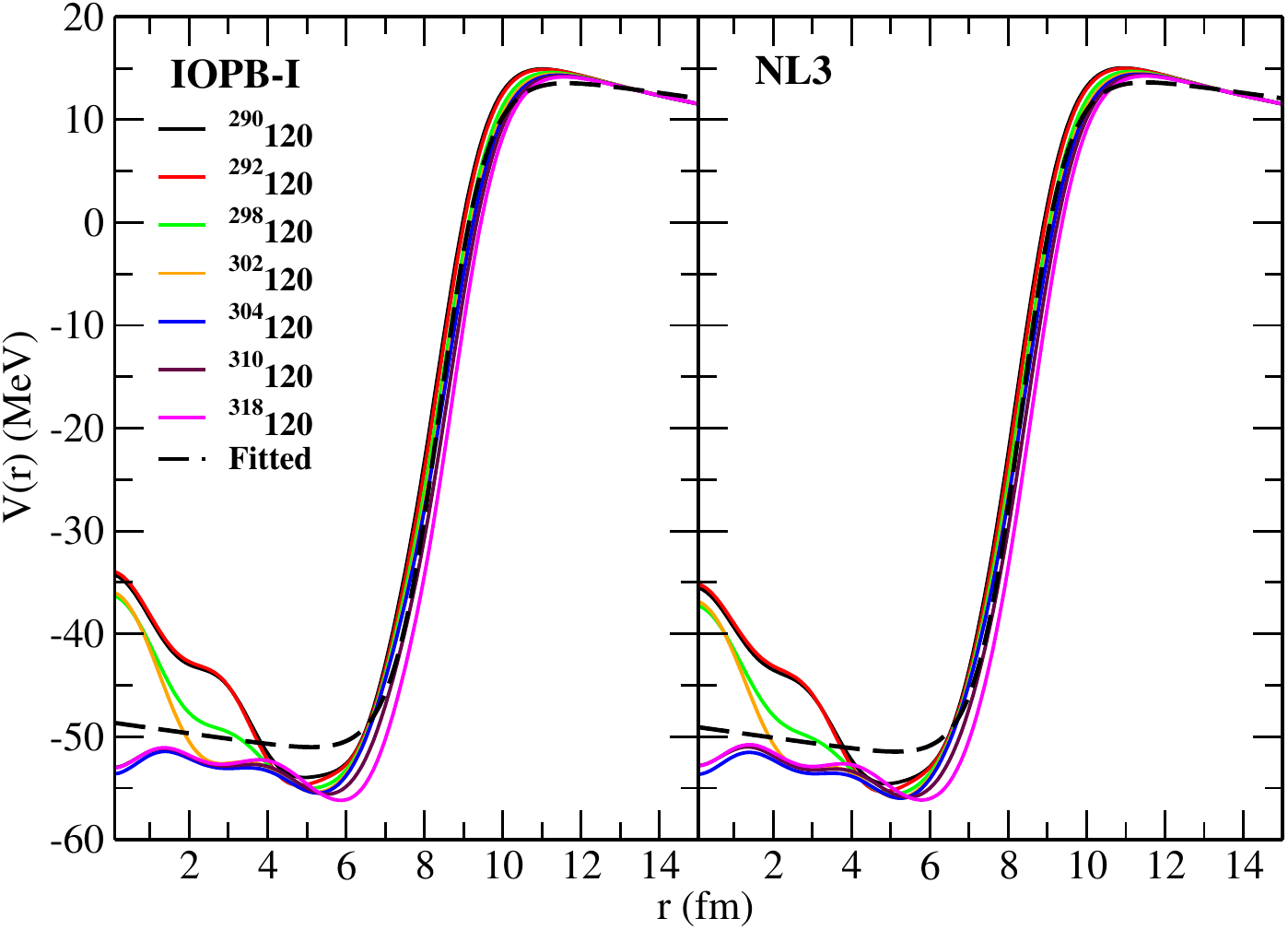}
\caption{The nuclear potential $V(r)$  for $^{290,292,298,302,304,310,318}$120 with IOPB-I and NL3 parameter sets. The dashed line is the fitting potential termed as $``SOA"$ potential. } 
\label{fig3}
\end{figure}

\begin{figure}
\centering
\includegraphics[width=0.8\columnwidth]{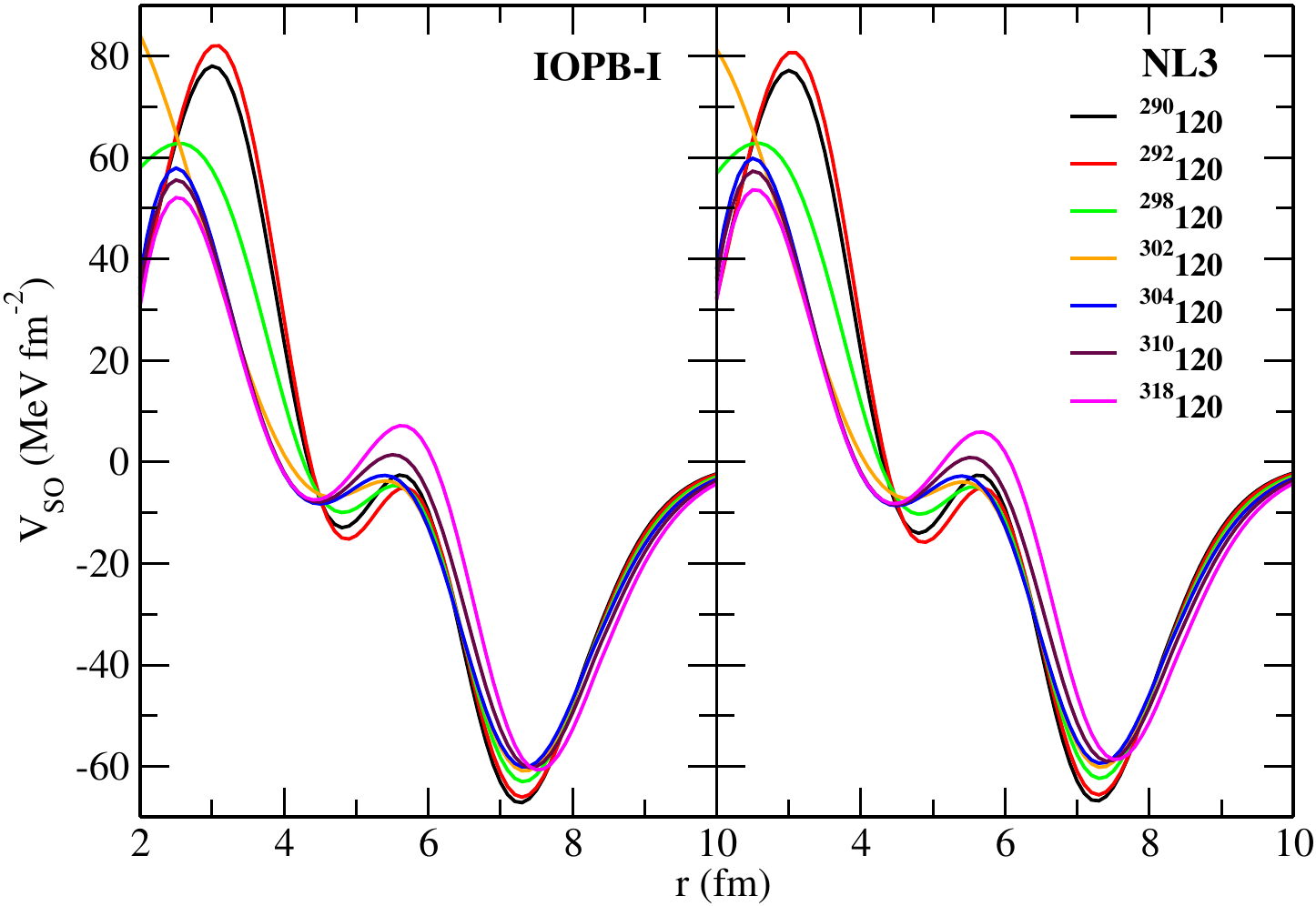}
\caption{The spin-orbit potential $V_{SO}(r)$ for $^{290,292,298,302,304,310,318}$120 with IOPB-I and NL3 parameter sets.  } 
\label{fig5}
\end{figure}

\begin{figure}
\centering
\includegraphics[width=1.0\columnwidth]{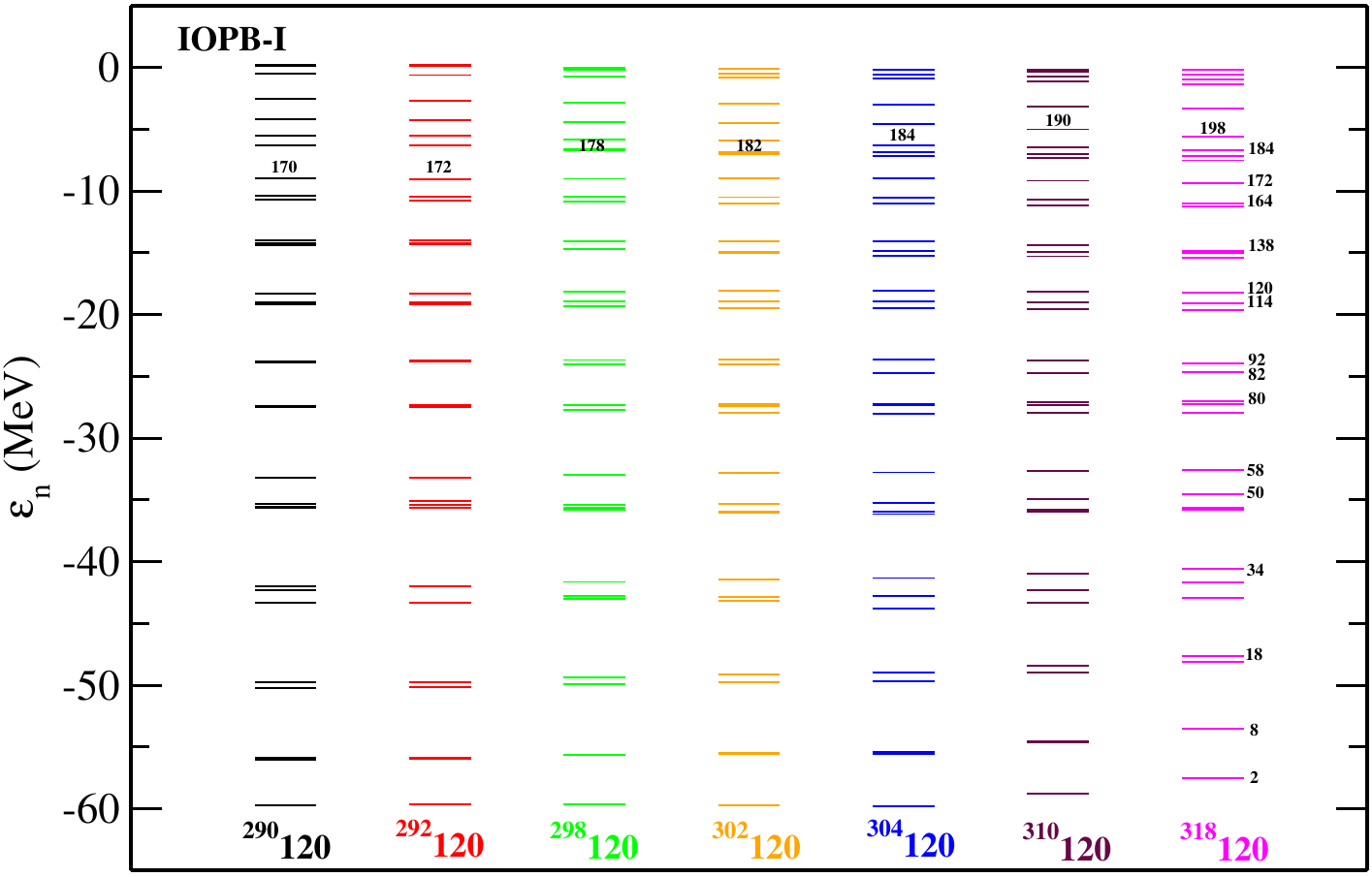}
\includegraphics[width=1.0\columnwidth]{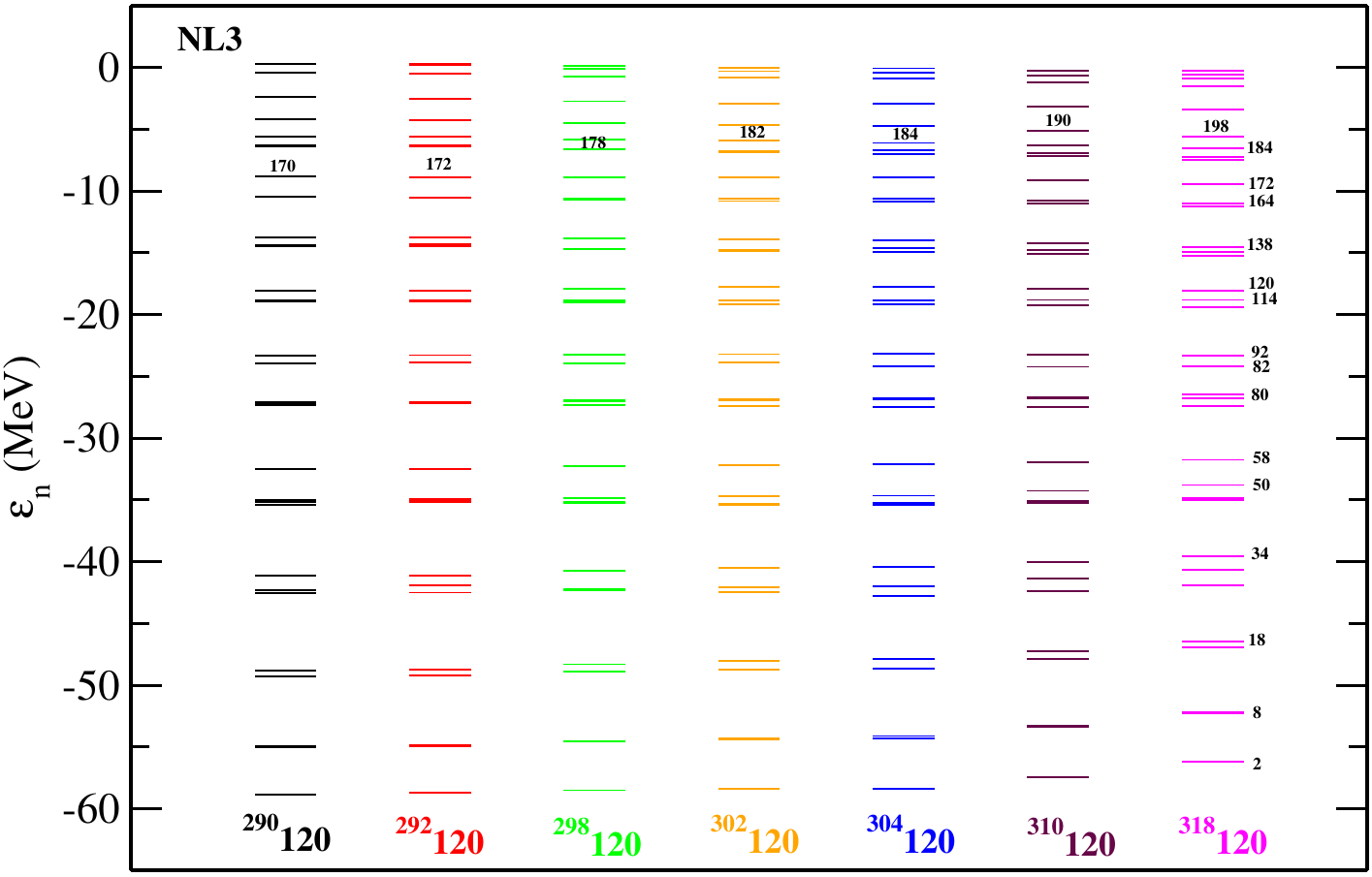}
\caption{ The single-particle levels $\epsilon_n$ of neutron for $^{290,292,298,302,304,310,318}$120 with IOPB-I and NL3 parameter sets. } 
\label{fig6}
\end{figure}

\begin{figure}
\centering
\includegraphics[width=0.93\columnwidth]{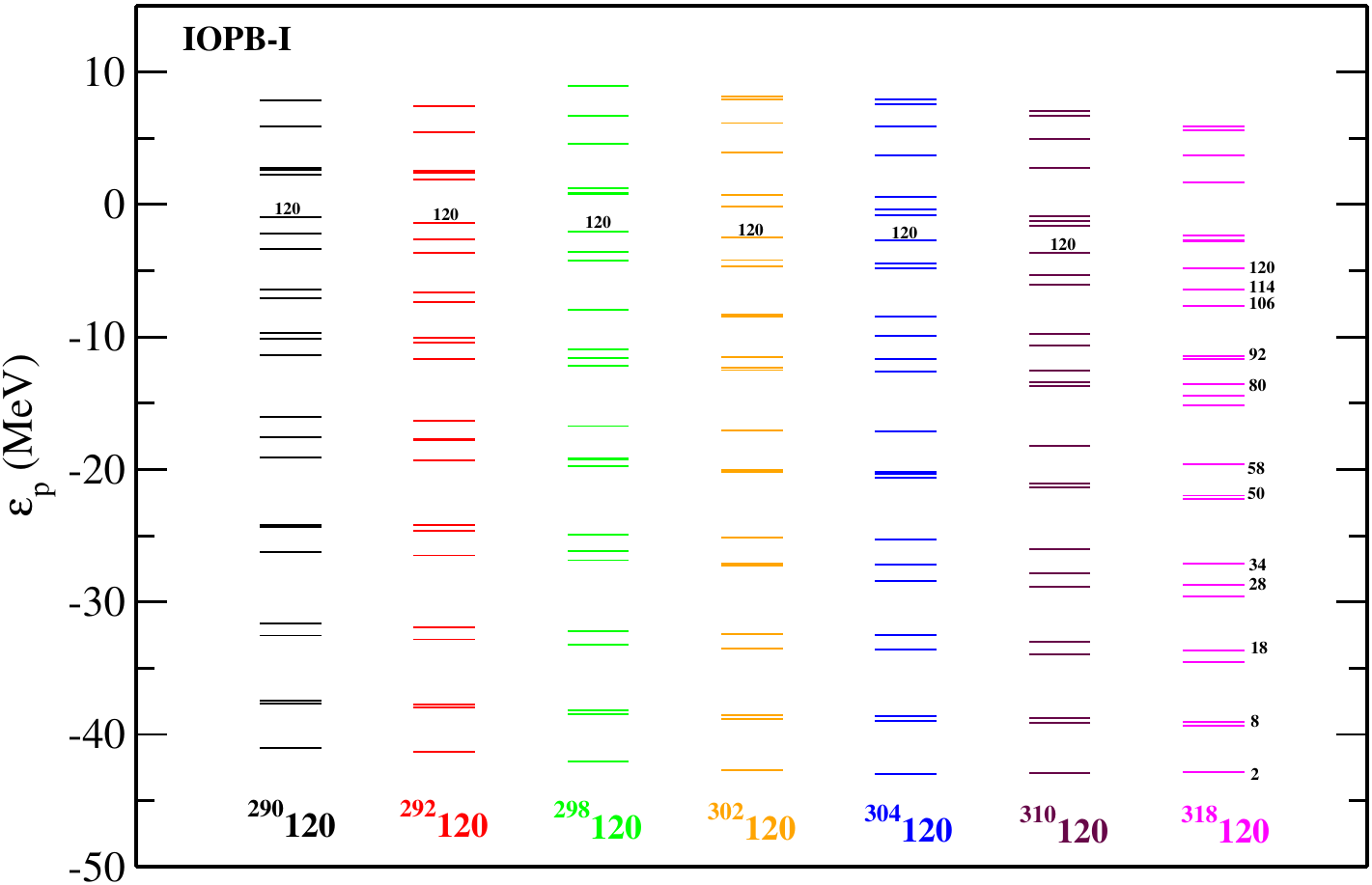}
\caption{ The single-particle levels $\epsilon_p$ of proton for $^{290,292,298,302,304,310,318}$120 with IOPB-I parameter set. The celebrity Z = 120 shell gap (proton magic number) is distinctly visible in the spectrum. } 
\label{fig7}
\end{figure}

In this section, we present the results of our numerical calculations along with a discussion for the nuclear density and potential which has a connection with the nucleons distribution in various energy levels. 
\begin{table*}
\centering
\caption{The calculated binding energies (BE in MeV) of IOPB-I and NL3 forces are compared with the available experimental data \cite{wang2017}.}
\renewcommand{\tabcolsep}{0.67cm}
\renewcommand{\arraystretch}{1.14}
\begin{tabular}{cccccccccccccccc}
\hline \hline
Nucleus	&	IOPB-I  &	NL3 & Expt	\\
\hline
$^{256}$Fm	& 1883.90 & 1888.26 &   1902.53 \\
$^{257}$Fm	& 1889.95 & 1894.42 &	1907.50 \\
$^{257}$Md	& 1889.18 & 1893.29 &	1906.31 \\
$^{258}$Md	& 1895.35 & 1899.59 &	1911.69 \\
$^{256}$No	& 1881.62 & 1885.35 &	1898.63 \\
$^{257}$No	& 1887.99 & 1891.78 &	1904.28 \\
$^{259}$No	& 1900.51 & 1904.54 &	1916.59 \\
$^{256}$Lr	& 1879.49 & 1883.01 &	1893.93 \\
$^{256}$Rf	& 1876.45 & 1879.98 &	1890.67 \\
$^{257}$Rf	& 1883.61 & 1886.99 &	1897.09 \\
$^{258}$Rf	& 1890.71 & 1893.88 &	1904.69 \\
$^{259}$Db	& 1894.88 & 1897.94 &	1906.33 \\
$^{260}$Sg	& 1898.79 & 1901.85 &	1909.06 \\
$^{261}$Sg	& 1905.52 & 1908.76 &	1915.67 \\
$^{262}$Sg	& 1912.21 & 1915.60	& 	1923.39 \\
$^{264}$Hs	& 1915.14 & 1918.99 &	1926.77 \\
$^{265}$Hs	& 1922.38 & 1926.34 &	1933.50 \\
$^{266}$Hs	& 1929.53 & 1933.56 &	1941.33 \\
$^{269}$Ds	& 1940.14 & 1944.67 &	1950.29 \\
$^{270}$Ds	& 1947.71 & 1952.26 &	1958.51 \\                 
\hline
\hline
\end{tabular}
\label{tab1}
\end{table*}
The E-RMF results are depicted in Tables \ref{tab1} - \ref{tab2} and Figures \ref{fig1} - \ref{fig7} for binding energy (BE), charge distribution radius ($\rm R_{\rm ch}$), single-particle energy for nucleons configuration and density distribution of protons and neutrons along with the nuclear potential for some selected superheavy isotopes for Z = 120. The spin-orbit interaction as a function of radial coordinate is also given, as it is responsible for splitting the total angular momentum $J$ to $J=l+1/2$ and $J=l-1/2$ levels, where $l$ is the orbital angular momentum. Before discussing our findings, first of all, we want to justify our results by comparing them with the experimental quantities. For this, we have displayed the binding energies (BE) in Table \ref{tab1} for some of the heaviest nuclei of this region and compared with the available experimental data \cite{wang2017}. From Table \ref{tab1}, one can see that the computed BE matches well with the experimental values. For example, the IOPB-I and NL3 both predict the BE as 1947.7 and 1952.3 MeV for $^{270}\rm Ds$ with the experimental value of 1958.5 MeV. These predicted binding energies of NL3 and IOPB-I are quite close to the experimental data with an error of $0.55 \%$ and $0.31 \%$ with IOPB-I and NL3 sets, respectively. This accuracy persists valid for the entire region of the superheavy valley, which can be seen in Table \ref{tab1}.  Thus, our observation gives us confidence to extend the calculations for the Z = 120 isotopes. The results for binding energy, charge radius, neutron and proton chemical potentials and the single-particle energy of the last occupied and the first unoccupied orbits are depicted in Table \ref{tab2}. The most trusted quantity in nuclear property is the binding energies found to be almost equal with the NL3 and IOPB-I forces, although the origin of both forces is very different from one another. The binding energy arises as a direct consequence of single-particle energy, establishing the reliability of the single-particle energy values presented in this study. Consequently, the energy gap obtained in our analysis represents the most accurate and physically consistent result. The binding energy is determined by the delicate cancellation of substantial scalar and vector contributions; even a minor deviation in these components can lead to significant discrepancies in the binding energy of a nucleus. Given that our calculated binding energy exhibits excellent agreement with experimental data, we can confidently assert the authenticity of the single-particle energy levels. Similar to the BE, the $\rm R_{\rm ch}$ are also matching with each others. 
\begin{table*}
\centering
\caption{The binding energy (BE in MeV), charge radius $\rm R_{\rm ch}$ (in fm), neutron and proton chemical potential ($ \lambda_{n}$ \& $\lambda_p$) (in MeV) and $\epsilon_{n}$ (in MeV) for both last occupied \& unoccupied orbitals are given for IOPB-I and NL3 forces.}
\renewcommand{\tabcolsep}{0.2cm}
\renewcommand{\arraystretch}{1.14}
\begin{tabular}{cccccccccccccccc}
\hline \hline
Nucleus &Force& BE & $\rm R_{\rm ch}$ & $ \lambda_{n}$ & $\lambda_p$ & $\epsilon_{n}$ (Last occupied)  & $\epsilon_{n}$ (Last unoccupied)\\
\hline
\hline
$^{290}$120   
           & IOPB-I       & 2040.10 & 6.286 & -8.79 & -0.76 &-8.97 &-8.97  \\
           & NL3        & 2045.69 & 6.267 & -8.64 &1.72 & -8.82& -8.82 \\
           
\hline
$^{292}$120   
            & IOPB-I      & 2057.57 & 6.299 & -6.63 & 1.54 & -9.07 & -6.32 \\
            & NL3       & 2062.84 & 6.280 & -6.68 &1.42 & -8.91&-6.39  \\
           
\hline
$^{298}$120   
            & IOPB-I       & 2097.13 & 6.314 & -6.63 &0.52 & -6.64&-6.64  \\
            & NL3        & 2102.36 & 6.297 &-6.57 &0.53 & -6.61 & -6.61  \\
           
\hline
$^{302}$120   
            & IOPB-I      & 2123.50 & 6.325 & -6.60 & -0.37& -6.87 & -5.94 \\
            & NL3       & 2128.30 & 6.308 & -6.45 & -0.27 & -6.76 & -5.94 \\
           
\hline
$^{304}$120   
            & IOPB-I      & 2135.69 & 6.329 & -6.22 &-0.94 & -6.33 & -4.56 \\
            & NL3        & 2140.28 & 6.312 & -5.99 & -0.79 & -6.11 & -4.71 \\
           
\hline
$^{310}$120  
            & IOPB-I         & 2165.68 & 6.386 & -5.06 & -1.80 & -5.01 & -5.01\\
             & NL3        & 2170.97 & 6.371 & -5.16 &-1.66 & -5.11&-5.11  \\
                      
\hline
$^{318}$120  
            & IOPB-I     & 2206.38 & 6.460 & -3.66 & -3.19 &-5.58& -3.32\\
            & NL3        & 2212.18 & 6.447  & -3.75  &-3.13 & -5.61& -3.39 \\
\hline
\hline
\end{tabular}
\label{tab2}
\end{table*}
In Figure \ref{fig1}, we have shown the neutron and proton density distributions for some of the selected isotopes of Z = 120 nucleus. In this figure, the proton number Z = 120 is kept fixed with a variation of neutron number in the isotopic series. The densities are also compared with the NL3 calculations. In both the cases, we find almost identical proton and neutron distribution confirming the model independent of the outcome. A further inspection shows that the presence of neutrons inside the nucleus is significantly different in an isotopic chain, which is seen from the densities in the region $r = 0$ to 4 fm. The nuclei $^{290,292,298,302}$120 show a bubble-like structure near the center, contrary to the maximum densities for $^{304,310,318}$120 \cite{shailesh13}. 
The presence of a bubble or semi-bubble structure can be characterized using the depletion fraction (D.F), as defined in Refs. \cite{gras09,gras09a}. This quantity, expressed as a percentage, is given by:  
\begin{eqnarray}
(D.F)_{total} = \frac{(\rho_{max})_{total}-(\rho_{central})_{total}}{(\rho_{max})_{total}} \times 100.
\label{dfeqn}
\end{eqnarray}
In this expression, $\rho_{max}$ represents the maximum neutron density, while $\rho_{central} = \rho(r = 0)$ denotes the central neutron density. The subscript 'total' refers to the total density distribution, which includes both neutron and proton contributions \cite{co18,shuk11,shar15}. This change in the neutron distributions has a larger consequence on the nuclear potential, which makes a different magic number sequence than the valley of stability. On the other hand, because of the increase in N number, the arrangement of the fixed protons (Z = 120) remains unaffected except a marginal reshuffling in the core region. The proton distribution density for these nuclei is shown with the dashed lines in the lower portion of Figure \ref{fig1}. We do not see much difference in the density pattern either with the parameterizations or with the change in the neutron number in the isotopic chain.\\


Figure \ref{fig2} shows the total nuclear potential for $^{298}$Fl, $^{300}$Lv, $^{302}$Og and $^{304}$120 nuclei. Here, the neutron number is fixed (N = 184) and the atomic number changes with an unit of 2 to see the effect on the potential as a function of Z. The calculated potentials for both the forces (IOPB-I and NL3) are almost similar in magnitude and pattern. This means, the change in potential is mostly determined by the neutron numbers. This is a very important finding, which may have a larger effect on an asymmetric system, like neutron star structure. Similar to the effects of proton on nuclear potential, we have shown the nuclear potential as a function of neutron number in Figure \ref{fig3} for the Z = 120 isotopes. Unlike to the proton variant, we find huge influence of the neutrons in the core area of the nucleus. This nature is also evident from the density distribution analysis. We find a hill-type potential for some of the nuclei ($^{290,292,298,302}$120) as shown in the figure. \\

To study the structure of the potential, we fitted the potential $V(r)$ with different functionals. The obtained $V(r)$ shape is defined as:
\begin{equation}
V(r) = a e^{-b r} + \frac{c}{1 + e^{-d (r - e)}} + f r + g,
\end{equation}
\label{poteq}
where $a$, $b$, $c$, $d$, $e$, $f$, and $g$ are the fitted constants defined as 
$a = -35.733122$, $b = 0.000003$, $c = 69.129224$, $d = 1.684950$, $e = 8.317300$, $f = -0.535011$, $g = -13.293379$. The fitted potential is a combination of a few established functions. The short-range region is the Yukawa-like potential $V_1=a e^{-b r}$, the intermediate part is the Woods-Saxon-like shape $V_2=\frac{c}{1 + e^{-d (r - e)}}$. This part decides the central attractive nuclear potential. The long-range part is the linear combination of $fr$ and $g$ define as $V_3=fr+g$. The total nuclear potential for the superheavy region V is written as $V(r)=V_1+V_2+V_3$, is termed as $``SOA"$ potential. The fitting potential ($``SOA"$-potential) deviates from the actual nuclear potential by $\Delta=1.56$ is obtained by using the relation $\Delta = \frac{1}{N} \sum_{i=1}^{N} \left| V_{\text{data}, i} - V_{\text{fit}, i} \right|$, where N is the total number of data points. This provides an absolute measure of fitting accuracy, irrespective of relative scaling effects. The spin-orbit splitting can not be isolated from the $SOA-$potential, as the Eq. (\ref{poteq}) is obtained from an average fitting of the E-RMF potentials. The spin-orbit potential $V_{SO}(r)$ is obtained by transforming the E-RMF Dirac equation to it's Schr\"odinger equivalent form, by using the Foldy-Wouthuysen transformation \cite{foldy50}, which is written as $V_{SO}(r)=\frac{1}{2Mr^2} \frac{d}{dr}[U_v(r)+U_s(r)]$. The $U_V(r)$ and $U_S(r)$ are the total vector and scalar parts of the relativistic nuclear potentials, respectively, obtained from the E-RMF Lagrangian Eq. (\ref{eq1}). \\

Now the spin-orbit potential $V_{SO}$, which plays a major role in the magic number sequence is given in Figure \ref{fig5}. We get a maximum value in the central region of the nucleus ($r=3-4$ fm) and the minimum $V_{SO}$ is near the surface ($r=7-8$ fm). In addition to these two extremes, i.e., minimum $V_{SO}$ and maximum $V_{SO}$, we notice an intermediate rise at $\sim 5$ fm. This observation is true for both the parameter sets and also applicable for all the Z = 120 isotopes considered in the present report. The magic number sequence or the shell closure sequence is determined by the nuclear potential as well as the spin-orbit interaction. However, the $V_{SO}$ part is hidden in the total nuclear potential and does not reflect in the $``SOA"$ potential, but the effect is very much visible in the single-particle spectrum, which is shown in Figure \ref{fig6} for the arrangement of neutrons in various shells. The single-particle levels for some of the representative nuclei $^{290,292,298,302,304,310,318}$120
are plotted for IOPB-I and NL3 forces.
We find shell closure at neutron number 2, 8, 18, 34, 50, 58, 80, 82, 92, 114, 120, 138, 164, 172, 184 and 198 \cite{mehta02}. The orbits are filled sequentially as shown in the figure as $1(s_{1/2})^2$, $1(p_{3/2})^4 1(p_{1/2})^2$, $1(d_{5/2})^6  1(d_{3/2})^4$, $2(s_{1/2})^2 1(f_{7/2})^8 1(f_{5/2})^6$, $2(p_{3/2})^4 2(p_{1/2})^2 
1(g_{9/2})^{10}$, $1(g_{7/2})^8$, $2(d_{5/2})^6 2(d_{3/2})^4 1(h_{11/2})^{12}$, $3(s_{1/2})^2$, $1(h_{9/2})^{10}$, $2(f_{7/2})^8 1(i_{13/2})^{14}$, $2(f_{5/2})^6, 
3(p_{3/2})^4 3(p_{1/2})^2 1(i_{11/2})^{12}$, $2(g_{9/2})^{10} 1(j_{15/2})^{16}$, $2(g_{7/2})^8$, $3(d_{5/2})^6 3(d_{3/2})^4 4(s_{1/2})^2$, $1(j_{13/2})^{14}$. After investigating, we find large shell gaps at 2, 8, 18, 34, 50, 58, 80, 92, 120, 138, 164, 172, 184 and 198. This means, although there is shell closure at 82 and 114, we do not see substantial shell gaps in these two numbers to satisfy as magic numbers, because their shell gaps are very small in both the IOPB-I and NL3 parameter sets. Thus, some of the existing magic number like N = 82 washed away and some new gaps appear, such as N = 58, 80, 92, 120, 138, 164, 172, 184, and 198, which may be termed as a new sequence of shell gaps \cite{mehta02}. The existence of these combinations of nuclei can be verified experimentally. The analysis of the proton single-particle energy levels is shown in Figure \ref{fig7}. Similar to the neutron sequence, here the close shell pattern is 2, 8, 18, 28, 34, 50, 58, 80, 92, 106, 114 and 120. Out of these close shell proton numbers, the shell gaps at 106 and 114 are considerably small and can be overlooked while considering the magic number sequence. Thus, the proton magic numbers in the superheavy valley can be defined as Z = 2, 8, 18, 28, 34, 50, 58, 80, 92 and 120. In this case also, some of the known magic numbers disappear while the proton numbers 34, 58, 80 and 120 can be considered as the magic numbers in the superheavy region.  

\begin{figure}
\centering
\includegraphics[width=0.8\columnwidth]{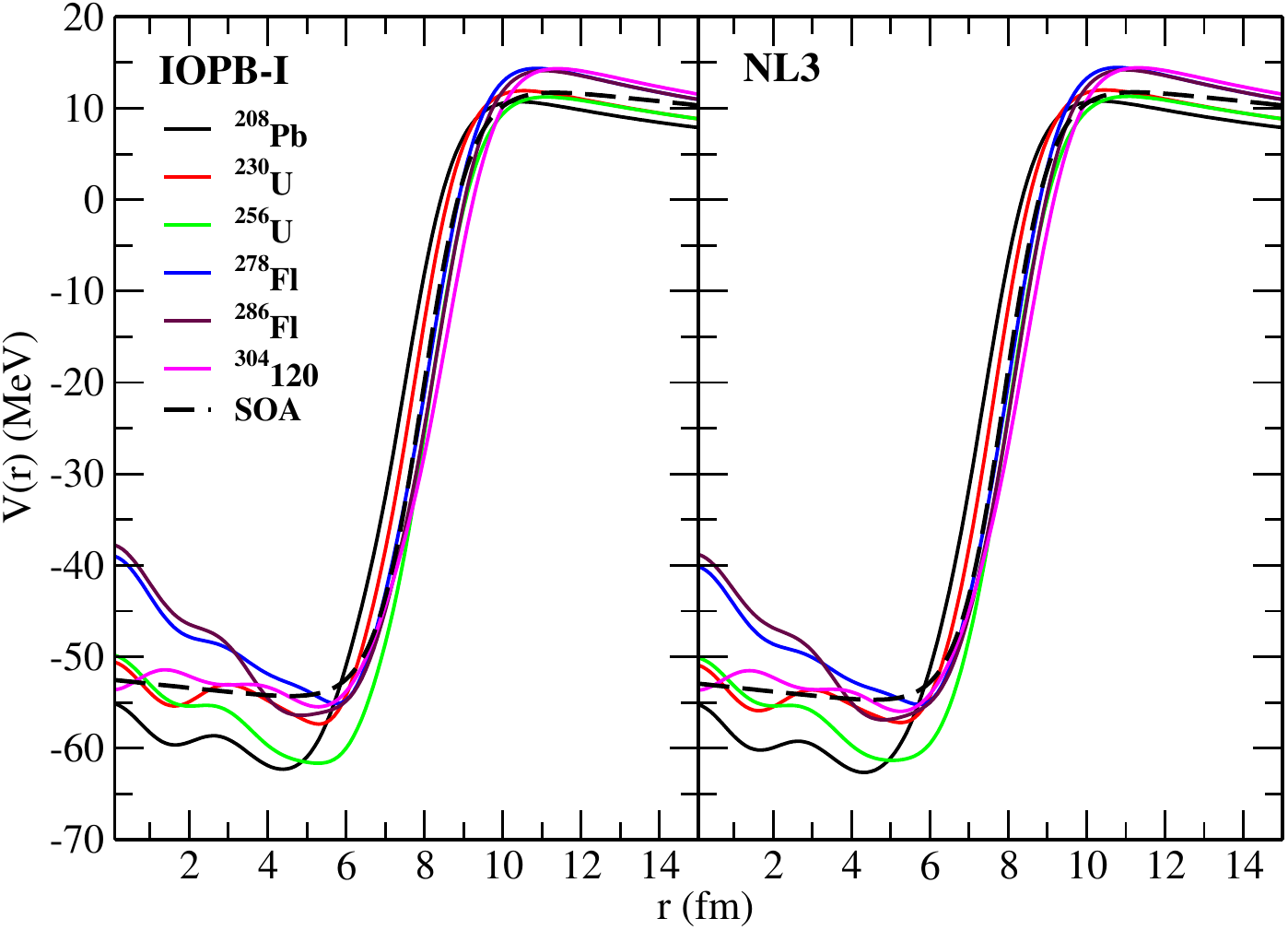}
\caption{ The nuclear potential $V(r)$  for $^{208}$Pb, $^{230}$U, $^{256}$U, $^{278}$Fl, $^{286}$Fl and $^{304}$120 with IOPB-I and NL3 parameter sets. The dashed curve is the $SOA-$potential obtained from Eq. (\ref{poteq}).} 
\label{fig4}
\end{figure}

After getting the final magic number sequence for neutrons and protons, we again demonstrate the nuclear potential for such a magic number of protons and neutrons. The Figure \ref{fig4} shows the nuclear potential for the selected combinations of Z and N as  $^{208}$Pb, $^{230}$U, $^{256}$U, $^{278}$Fl, $^{286}$Fl and $^{304}$120. The total $V(r)$ of neutrons and protons for the double closed nuclei are of similar kind with a slight variation on the wall of the potentials, justifying the magic nature of the potentials. The $SOA$ potential is also overlapped on it (dashed line) to see the overall nature of the newly predicted double close nuclei. For comparison, the potential of the known heaviest double close nucleus $^{208}$Pb is given (solid black line) in the plot. In general, we get a $SOA-$like potential for all the proposed double close nuclei. The structure of the central part of the potential deviates from the shape of the $SOA-$potential, showing the shell structure of the superheavy nuclei. However, the variation in shell effect is missing in the $SOA-$potential and necessary modifications are in the process.   

\section{SUMMARY AND CONCLUSIONS}
In summary, we calculate the binding energy, charge radius, nuclear density distributions, nuclear potential, spin-orbit interactions and the single-particle energy levels for some superheavy nuclei. From the analysis of nuclear potential, we found that the potential is almost in-sensitive to proton numbers, but very much influenced by the addition of neutron numbers. Thus, we find very different potential from one another with the addition of neutron numbers in the Z = 120 isotopic series. To see the shape of the potential, we performed a fitting exercise and find an analytic expression of the nuclear potential termed as $``SOA"$ potential. The spin-orbit interaction is hidden in it and can not be separated without the Foldy-Wouthuysen transformation \cite{foldy50}. Thus, the back calculations with the $``SOA"$ potential can not reproduce the magic number sequence.  The spin-orbit potential is also showed up and down with the radius of the nucleus. We find maximum spin-orbit at $3-4$ fm and minimum at $ 7 - 8$ fm for the considered mass region of the superheavy elements. In addition to these, we also noticed an up at $\sim 5$ fm.   

Taking into account the spin-orbit interaction and the structure of the potential, we find a different magic number sequence in the superheavy valley. Some of the well-known magic numbers are washed out and a few more new sequences have appeared due to the shape of the potential and the large spin-orbit interactions. Hence, the new magic number sequence for proton and neutron in the superheavy valley is found out to be Z = 2, 8, 18, 28, 34, 50, 58, 80, 92, 120 and 2, 8, 18, 34, 50, 58, 80, 92, 120, 138, 164, 172, 184, 198 respectively. We have also proposed some double closed configurations like $^{230}$U, $^{256}$U, $^{278}$Fl, $^{286}$Fl and $^{304}$120, which may be synthesized experimentally.   
\section{ACKNOWLEDGEMENT:}
\noindent 
JAP acknowledges IOP for providing computer facilities and support from Science and Engineering Research Board (SERB) through the contingency of Ramanujan Fellowship File No. RJF/2022/000140.  \\

\end{document}